# Sad syntax? Tonal closure Affects Children's Perception of Emotional Valence

## Assaf Suberry[1,2*], Neta B. Maimon[3] and Zohar Eitan[4]

[1] The Interdisciplinary Department of Social Sciences , Bar Ilan University, POB 9000, Ramat Gan 5290002, Israel

[2] The Faculty of Music Education, Levinsky College of Education, POB 48130, Tel Aviv 15, Israel

[3] The School of Psychological Sciences, Tel Aviv University, POB 39040, Tel Aviv 69978, Israel

[4] Buchman-Mehta School of Music, Tel Aviv University, POB 39040, Tel Aviv 69978, Israel

* To whom correspondence should be addressed. E-mail: assaf.su@gmail.com

**Abstract**

Western music is largely governed by tonality, a quasi-syntactic system regulating musical continuity and closure. Converging measures have established the psychological reality of tonality as a cognitive schema raising distinct expectancies for both adults and children. However, while tonal expectations were associated with emotion in adults, little is known about tonality's emotional effects in children. Here we examine whether children associate levels of tonal closure with emotional valence, whether such associations are age-dependent, and how they interact with other musical dimensions. 52 children, aged 7, 11, listened to chord progressions implying closure followed by a probe tone. Probes could realize closure (tonic note), violate it mildly (unstable diatonic note) or extremely (out-of-key note). Three timbres (piano, guitar, woodwinds) and three pitch heights were used for each closure level. Stimuli were described to participants as exchanges between two children (chords, probe); participants chose one of two emojis, suggesting positive/negative emotions, as representing the "2nd child's" response. A significant effect of tonal closure was found, with no interactions with age, instrument, or pitch height. Results suggest that tonality, an abstract, non-referential cognitive schema, affects children's perception of emotion in music early, robustly and independently of basic musical dimensions.

# Introduction

**Tonal cognition and emotion**

Generally, the term "tonality" denotes a structure of interrelationships among pitches within a musical context, in particular their relationship with a central, maximally stable pitch (the tonic). In a more restricted sense, the term refers to the interrelationships among tones or chords in music utilizing Western keys (major/minor), which comprise most Western music ("classical" and popular) since the 17th century. In these contexts, the tonic note (scale degree 1) and, to a lesser extent, the other members of the tonic triad (scale degrees 3 and 5) are considered most stable and closural; the other diatonic scale degrees (2, 4, 6 and 7) are less stable and imply continuation to their more stable neighbors, 1, 3, or 5; the remaining five chromatic ("out of key") tones are the least stable, and strongly imply continuation to adjacent diatonic (within-key) notes.

The psychological reality of tonality as a cognitive schema orienting the listener has been strongly established empirically (see Shanahan, 2017; Krumhansl, 2004 for research surveys). Studies applying converging experimental paradigms – explicit measurements, such as sung continuations and goodness of fit ratings (e.g., Carlsen, 1981; Krumhansl, 1990; Cuddy & Lunney, 1995), as well as implicit ones, such as musical priming and event-related potentials (ERP; e.g., Tillman & Bigand, 2004; Granot & Donchin, 2002) – have suggested that listeners implicitly abstract a tonal hierarchy, and the sets of melodic and harmonic expectancies it entails, closely matching those conjectured by music theorists' models (e.g., Rameau, 1722; Fétis, 1844; Schoenberg, 1978).

As a cognitive schema, tonality generates a set of implicit or explicit expectancies. Thus, within a tonal context, some continuations (e.g., B→C in C major) would be more likely than others (e.g., B→C♯ in C major), and hence more strongly expected. Such expectations may be based on implicit statistical learning, produced through exposure to tonal music, rather than on explicit rule-based learning (Huron, 2006); hence, neither musical training nor explicit command of musical rules and concepts are prerequisites for their emergence (see, e.g., Tillmann, 2005, for related empirical work).

The interaction of expectations with incoming information, which could realize, violate, or suspend such expectations, may generate emotional responses. Violations of expectations -- in music as well as in other domains -- have been associated with a range of emotions, from generalized, unspecified arousal to specific emotions such as surprise, tension, fear, anger, or anxiety (Huron, 2006; Miceli & Castelfranchi, 2014; Proulx, Inzlicht, & Harmon-Jones, 2012). In musical contexts, composers' or performers' manipulations of schematic expectations, and listeners' responses to such manipulations, have long been proposed as central determinants of emotion – both felt emotions, evoked in the listener, and emotions perceived by listeners as expressed by the music (Huron, 2006; Juslin, Harmat, & Eerola, 2014; Juslin & Västfjäll, 2008; Meyer, 1956).

A number of empirical studies, applying subjective reports, implicit behavioral paradigms, as well as physiological and neuro-physiological measures, indicate that violations of tonal expectations are associated with heightened emotions -- felt or perceived. In Sloboda (1991), listeners reported sensing shivers concurrently with the advent of unexpected harmonies. Steinbeis, Koelsch, & Sloboda (2006) presented to participants (musicians and nonmusicians) several versions of Bach chorales, manipulating the degree of expectancy of a single chord in each version. Participants reported elevated tension and overall emotionality for unexpected chords, compared to expected ones; physiological measures, particularly heightened skin conductance corresponded with these subjective reports (note, however, that skin conductance was associated elsewhere with positive music-related emotions; Lundqvist, 2009). Koelsch, Fritz, & Schlaug (2008) presented to nonmusicians chord progressions in which the final chord could be either expected (tonic resolution) or highly unexpected (a chromatic substitute). Participants rated unexpected progressions as less pleasant than expected ones. Correspondingly, fMRI data indicated increased amygdala activity associated with the unexpected progressions. Similarly, in Juslin et al. (2014), deviations from tonal expectancy were associated by participants with negatively-valenced emotions, particularly anxiety, anger and irritation. Recently, Maimon, Lami, and Eitan (2018) demonstrated that listeners systematically associate the degree of tonal stability with visually-depicted emotional valence. Applying a modification of the probe-tone paradigm (Krumhansl & Kessler, 1982), they presented participants with a tonal context, followed by a probe tone (one of the 12 chromatic tones in each trial), and asked them to match each probe with one of seven facial expressions, ranging gradually from sad to happy. Participants (musicians and nonmusicians) matched more stable tones (e.g., the tonic note) with happier faces, and unstable ones (e.g., chromatic notes) with sad faces, correlating tonal hierarchy with emotional valence. Comparable results were obtained in a verbal rating paradigm by Arthur (2018). Overall, these studies suggest that realizations of the schematic expectations generated by tonality are mostly associated with positive valence, while violations tend to be associated with strong, often negative valence. These associations were found for evoked emotions (emotions aroused by the stimuli), as indicated by both subjective responses and physiological measures, and perceived emotions (emotions recognized by participants as what the music expresses).

How do the emotional associations of tonality interact with those evoked by other musical features? Tonality is, of course, not the only aspect of music associated with emotion. Auditory dimensions such as pitch height, or musical timbre ("sound color") were shown to affect music-related emotions in fairly consistent ways (for research reviews see Gabrielsson, 2011; Gabrielsson & Lindström, 2010). Higher pitch, for instance, associates in musical contexts with positive emotional valence, particularly happiness, and low pitch – with sadness (Eitan & Timmers, 2010; Gabrielsson & Lindström, 2010). Such associations develop early: children aged 4-12, for instance, used a higher pitch register while singing with an intent of evoking happiness, and lower pitch when intending to evoke sadness (Adachi & Trehub, 1998). Musical timbre has also been systematically associated by listeners with emotional valence, arousal, and tension. These associations, involving spectral, temporal and spectro-temporal aspects of timbre (Eerola, Ferrer, & Alluri, 2012; Gabrielsson & Juslin, 1996), may also be affected by social tagging or other conventional connotations attributed to a recognizable musical sound (Ferrer & Eerola, 2011). To our knowledge, no empirical study has systematically examined how the emotional connotations of tonality interact with those of musical features such as pitch height or timbre, themselves associated with emotional expression. This study presents, within a developmental context, a new behavioral paradigm enabling such investigation.

**The development of children's tonal cognition**

The perception of different aspects of tonality, such as consonance/dissonance, scales, intervals, or harmonic progressions, matures gradually during different stages of children's development (for a review see Corrigall & Schellenberg, 2016). In the first year of life infants are sensitive to simple harmonic ratios, melodic contour and phrasing (Cohen, 2000; Trehub & Trainor, 1993), and prefer to listen to consonant over dissonant intervals (Trainor, Tsang & Chenung, 2002). Yet the apprehension and representation of tonal hierarchy per se emerge at later developmental stages, since discerning regularities among tones requires considerable exposure to music, as well as mature memory resources and attentional skills (Costa-Giomi, 2003; Sloboda, 1985; Zimmerman, 1993). Children begin to discern closure of tonal musical at the age of three (Kragness & Trainor, 2018), and may implicitly recognize notes as belonging to a musical scale by the age of four (Trainor & Trehub, 1994). Other tonality-specific skills, such as singing with tonal base or distinguishing between tonal and atonal music, may be discerned between the ages four and six (Dowling, 1999; Thompson, 2015 ). Sensitivity to harmonic progressions seems to appear between the ages five and seven (Speer & Meeks, 1985; Cuddy & Badertscher, 1987; Trainor & Trehub, 1994; Trainor & Corrigall, 2010; but see Corrigall & Trainor, 2019, an EEG study suggesting responses to harmonic violations by 3.5 year olds).

By age seven, children reliably show some degree of sensitivity to the principal tonal features (Krumhansl & Cuddy, 2010; Trehub & Weiss, 2017), and at 10 or 11, their tonal aptitudes largely approach adults' (Lamont & Cross, 1994). However, research concerning the development of tonal cognition during that age span (7-11) is equivocal. While no significant differences were found between six- and 11 years olds' sensitivity to tonal hierarchy deviations in both explicit (Cuddy & Badertscher, 1987; Speer & Meeks, 1985, both using the probe-tone method) and implicit (Schellenberg, Bigand, Poulin-Charrannot, Garnier, & Stevens, 2005) tonal priming tasks, other studies do show a developmental course during that period (e.g., Krumhansl and Keil, 1982; Wilson, Wales & Pattison, 1997). Lamont and Cross (1994), who applied different tonal primes within the probe-tone paradigm (also used in the present study) to children between six and 11 years old, conclude that while young children's representation of tonal pitch relationships is already remarkably stable, that representation is as yet abstracted and independent of temporal relationships. At later ages, sensitivity to temporal features of tonality increases, leading to its internalization as a time-dependent schema.

While there is ample evidence that children develop tonal expectations, little is known about the emotional associations of such expectations among children. As noted above, violations of tonal expectancies were associated with emotional responses in adults, regardless of formal musical training. Since young children also develop implicit tonal expectancies, one may hypothesize comparable effects among them. Indeed, as McPherson and Schubert (2016) suggest, children's perception of emotion in music, eralier (ages 3-7) dominated by referential processes (associated emotion in music with specific examples), is governed by age 7 by schematic, or "absolutist" processes, associating structural features in music (such as major vs. minor mode) with emotion. However, while the associations of mode with emotional valence have been studied in children (Gregory, Worrall & Sarge, 1996; Dalla Bella et al., 2001; Ziv & Goshen, 2006; Nieminan et al, 2012), hardly any empirical research has investigated whether tonal stability or tonal expectancies are associated with emotion among children, and how such associations may develop. Some studies examining children's tonal perception used "happy" and "sad" icons as response aids (Speer & Meeks, 1985; Cuddy & Badertscher, 1987; Lamont & Cross, 1994). Yet these icons merely served to gauge "goodness of fit"– i.e., indicate to what degree probe tones match their tonal context -- rather than evaluate participants' perceived or felt emotion. The present study aims to address this lacuna, systematically investigating how children associate tonal stability with emotional valence, and whether such associations differ between younger (seven years old) and older (11 years old) children.

**Overview of the present study**

This study, then, aims to address two understudied issues concerning tonal perception and its emotional connotations. First, we investigate whether children of two age groups (7, 11) systematically associate tonal stability with emotional valence, thus examining whether children's implicit perception of tonality also engenders emotional connotations, as in adults, and whether such connotations differ for younger and older children. Second, we examine whether the emotional connotations of tonality interact with those of other musical features (pitch height and instrumental timbre).

In a preliminary investigation (Pretest 1) we replicated with 7 and 11 years old children a recent study which examined how adult listeners associate tonal scale-degrees with emotional valence, as depicted by facial expression (Maimon et al., 2018), via an adaptation of the probe tone paradigm (e.g., Krumhansl and Kessler, 1982). Unlike the adults' experiment, where results suggest systematic relationships between tonal stability and valence, this pretest revealed no such relationships among children. Hence, in the main experiment we introduced a new design, adapted to young children. That design also enabled a systematic examination of how tonal stability interacts with other musical features (pitch height and instrumental timbre) in the representation of emotion.

To facilitate the task and adapt it to children, we accommodated it within a familiar social context (a conversation between two children), and reduced information load and complexity (e.g., replacing graded response options with a binary response – selecting one of two contrasting emojis). To examine the interaction of tonal stability with other musical features, we produced a matrix of musical stimuli in which each level of tonal stability was systematically associated with several different pitch heights and several instrumental timbres. As in Pretest 1, participants were children of two age groups, seven and 11 years old. Results demonstrated robust effects of tonal stability on perceived emotion (happy emojis were chosen more often for tonally stable stimuli, and sad emojis – for unstable stimuli), with no interaction with other musical dimensions or with age.

## Pretest 1[1]

Pretest 1 replicated Experiment 1 of Maimon et al. (2018, submitted), which examined how adult listeners associate tonal scale degrees with emotional valence, as depicted by facial expression. Adapting the standard probe-tone paradigm often used to gauge tonal perception (e.g., Krumhansl and Kessler, 1982), 7 and 11 year-olds heard a key-establishing context (e.g., a cadence) followed in. each trial by one of the 12 chromatic scale degrees. They were presented with a series of emotional faces (ranging from sad to happy), and asked to choose the face which (according to their subjective view) best matched the expression evoked by the probe tone. In contrast to the adults' experiment which served as a model for that pretest, where results suggested systematic relationships between tonal stability and valence (more stable probes were associated with happier faces), the pretest revealed no such relationships among children.

These results may suggest that children, unlike adults, do not yet associate tonal stability with emotional valence. Alternatively, the exact replication of stimuli and procedure originally created for adult participants could hinder children's performance, a different experimental design, more appropriate for young children, may reveal higher sensitivity to the emotional connotations of tonal stability. Both children's informal post-experiment responses and the need, often emphasized by developmental studies (e.g., Ilari & Young, 2016; Christensen & James, 2008, p. 3), to generate higher involvement and active participation of children participants, suggest that an experimental design adapted for children could better reflect their abilities and sensitivities.

Several aspects of the pretest could hinder children's performance. First, both visual and auditory stimuli included a large number of gradations (seven facial gradations in four versions,

---

[1] For a complete report of Pretest 1, see Appendix.

12 probe tones), a fact which could make the task difficult and confusing for children. The fairly large (96) number of trials could produce in a high information load, leading to fatigue and indifference. Indeed, several participants complained that "the tones all sounded the same" throughout the experiment. Furthermore, both visual stimuli (foreign-looking adult faces, with their extremities cut) and unfamiliar auditory timbre (Shepard tones) used in the pretest could further intimidate the children, elevating emotional distance in an experiment aiming to gauge perception of emotion. In designing the main experiment, we tried to address these issues.

## Main Experiment

In view of the results of Pretst 1, this experiment presents a new research paradigm, aimed to facilitate a simplified task suited to children. In addition, the new paradigm enables a systematic investigation of interactions between tonal stability and other musical features – pitch height and instrumental timbre. The main differences between the two experimental designs are as follows:

1. <u>Adapting the task to children:</u>

    a. **Graded vs. Binary response.** Pretest participants could be confused by the subtle gradations of the facial expressions presented to them. To facilitate participants' decisions, we introduced a binary response (positive or negative valence), represented by two contrasting emojis.

    b. **Introducing a familiar social context**. In lieu of the rather abstract instructions used in the pretest (see Appendix), we introduced a child-friendly task, asking the children to imagine a conversation between two kids (representing the tonal element and the probe tone), and select the emoji representing the response of the "2$^{nd}$ kid" (the probe tone).

c. **Using realistic instrumental timbre**. In lieu of the unfamiliar timbre (Shepard tones) used in the pretest, we used samples of several familiar musical instruments, aiming to generate greater interest and attention among the young participants.

*d.* **The tonal context: enhancing expectation**. The pretest (following Krumhansl & Kessler, 1982), used a single triad or a perfect authentic cadence (IV-V-I), followed by a probe. As perfect cadences end with complete tonal closure, strong expectations for any specific continuation could be attenuated. IN the main experiment. we used instead an incomplete cadence (IV-$I^4_6$-V7), strongly implying specific continuity and completion – a resolution to the tonic.

*e.* **Reducing the number of probe tones**. In the pretest, we examined the response to all 12 chromatic probe-tones. To reduce information load and introduce a sharper distinction between stimuli, the main experiment used 3 scale degrees only, each representing a different level of stability and closure: a stable diatonic note (the tonic), fully realizing the closure implication; an unstable diatonic note ($6^{th}$ scale-degree), mildly violating the closure implication mildly; or a chromatic (out-of-key), highly unstable note, strongly violating closure (raised $4^{\#th}$ scale-degree).

*f.* **Reducing session duration**. The main experiment lasted approximately 20 minutes, about 10 minutes shorter than the pretest. Session duration is a significant factor in behavioral studies involving children; as noted in earlier studies (e.g., Dalla Bella et al., 2001), reducing overall duration may enhance attention and reduce fatigue among participants.

a.

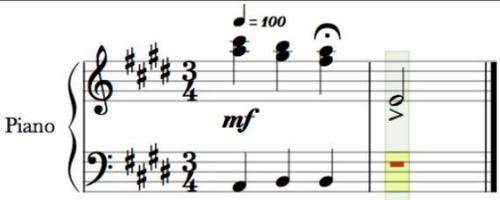

*Figure 1: Sample stimuli of the main experiment. A harmonic progression (IV-I$^4_6$-V$^7$) in three different keys, ending with a probe tone (instrumental timbre – piano). In all three examples, the same pitch (E4) serves as a probe tone, yet its relationship with the preceding tonal context is altered. In the top panel (E major key) the probe is the tonic note, which realizes the closure implication; in the middle panel (G major) the probe serves as the 6$^{th}$ scale degree, violating it mildly; in the lower panel (B-flat major) it serves as a chromatic (4$^{\#}$) degree, extremely violating closure implication.*

2. <u>Examining interactions with other musical variables</u>

To enable a systematic examination of the interaction of tonal stability and pitch height, we introduced a "reversed" probe-tone paradigm. In the standard probe-tone method (e.g., Krumhansl & Kessler, 1982), each block presents a single element, priming a single musical key. This element is repeated in all trials within the block, while probe tones differ for each trial. In contrast, in our Main Experiment the same probe tone is primed (in different trials) by different keys, so that it may function as different scale-degrees. For instance, in Figure 1 the same probe tone (E4) serves as the stable tonic in E major (top panel), a mildly unstable diatonic 6$^{th}$ degree in G major (middle panel), and a highly unstable, chromatic 4# in B-flat major (lower panel). E-flat4 (in the keys of E-flat, G-flat, and A major) and F4 (in the keys of F, A-flat, and B major) similarly presented all three functions. Furthermore, each cell in the above 3 (scale degrees) X 3 (probe tone) matrix is presented in 3 instrumental timbres (piano, guitar, woodwinds), thus enabling an examination of the independent effects and interactions of all three dimensions (tonal stability, pitch height, instrumental timbre).

## Method

Prior to the main experiment, we conducted a preliminary test (Pretest 2) with 13 participants (six seven years olds; seven 11 years olds), aiming to establish a paradigm that fits children and evaluate whether children would be able to understand and complete the task.

**Participants.** 56 children aged 7 and 11, who had not participated in either pretests, served as participants: 26 in the younger group (11 boys, 15 girls), 30 in the older group (15 boys, 15 girls). One participant of the younger group and three of the older group were excluded from the sample for failing to complete the experiment or for lack of cooperation (e.g. selecting the same emoji for all stimuli). Thus, the final sample included 25 children in the younger group, and 27 in the older group.

All participants in the older group had a history of 6 to 24 months of music classes at school; five of them have played a musical instrument for 6 to 36 months. None of the younger participants have taken music lessons, but two of them have played a musical instrument for 6 to 12 months. .

All children had everyday exposure to various genres of Western tonal music. All children spoke fluent Hebrew (the language in which the experiment was conducted).

Children were all sampled from the same after-school community. Participants' parents signed informed consent forms, and all children agreed to participate willingly and could quit at any time with no sanctions. Children received a small sweet prize as a compensation.

**Apparatus.** The stimuli were presented using a Dell N5110 computer on a 15.6 inches screen and a Dell E6420 computer on a 14 inches screen. Auditory stimuli were delivered through Behringer HPM1000 headphones. The experiment was programmed and performed using Netbeans. One or two participants sat in a quiet room in front of separate computers, distant from each other and without eye contact.

**Stimuli.** Auditory stimuli were generated using the Notion music notation software (version 6), with its electric guitar, piano, and woodwinds (two clarinets and a bassoon) samples, originally recorded by the London Symphony Orchestra (All auditory stimuli are available here: https://osf.io/te5da/).

On each trial, participants heard an incomplete cadence (IV-I$^4_6$-V$^7$), strongly implying resolution to a tonic (I), followed by a single probe tone. The cadence was comprised of a three-voice texture; to minimize short-term memory priming effects, the tonic note was omitted from the texture in the IV and I$^4_6$ chords. Probes could realize the closure implication (tonic note), violate it mildly (6$^{th}$ diatonic scale-degree) or extremely (a chromatic, out-of-key note, 4$^{\#th}$ scale-degree). The probe was one of three pitches: E-flat4, E4 or F4. To minimize response biases due to pitch proximity, the probes were more than an octave lower than the upper line of the chord progression. The duration of each chord was 600 ms, and of the probe tone -- 1200 ms. A brief pause (150 ms) was inserted between the cadence and the probe tone. Altogether, there were 27 different stimuli, comprising a matrix of 3 (instrumental timbres) x 3 (probe-tone pitch heights) x3 (probe-tone scale-degrees). For notated examples of stimuli, see Figure 1.

**Visual stimuli:** The visual stimuli consisted of two emojis (Figure 2). The two emojis were selected by children participating in Pretest 2 from a pull of four emojis (two of them positive and two of them negative) as best expressing positive (happy) and negative (sad/angry) emotions. The diameter of each emoji, as presented on the screen, was 1.96 inches.

**Procedure.** Participants first heard a recorded explanation of the task, while an identical printed text was simultaneously presented on the screen. They were instructed to imagine that the sounds heard in each trial are a conversation between two children. The 1$^{st}$ child says something, using a few tones (i.e. the cadence), and the 2$^{nd}$ child replies, using a single tone (i.e., the probe tone). The 2$^{nd}$ child can reply in one of two manners: a positive (happy) manner, represented by the "positive" emoji, or a negative (sad or angry) manner, represented by the "negative" emoji. A visual illustration (without sound) was then presented on the screen: "The 2$^{nd}$ child can reply in a positive manner (the positive emoji was marked by an animated cue), but he can also reply in sad or angry manner (the negative emoji was marked by an animation cue)." The participants were instructed to press, in each trial, the emoji that fits the 2$^{nd}$ child's response better. In each trial, children could listen to the musical stimulus as many times as they wished; if uncertain, they could press a designated icon at the bottom of the screen and listen again. The auditory and visual stimuli were presented simultaneously in each trial with the instruction: "Press the face depicting the response of the 'second child' (the last note)".

The experiment consisted of three blocks, each in a different instrumental timbre (piano, guitar, or woodwinds). Each block included 20 consecutive trials. The first two trials, using two randomly selected stimuli, served as practice trials. These were followed by 18 experimental trials (3 categories of tonal stability*3 probe tones, each presented twice). The stimuli were presented for each participant in a semi-random order (identical stimuli could not appear successively). Block order was counterbalanced across subjects. There was a one-minute pause between blocks. To encourage persistence, a picture of a virtual gold bag was shown during the pause; participants could accumulate three virtual gold bags in order.

In sum, the experiment consisted of 54 experimental stimuli (the 3x3x3 matrix described above, with each stimulus presented twice), and lasted approximately 20 minutes.

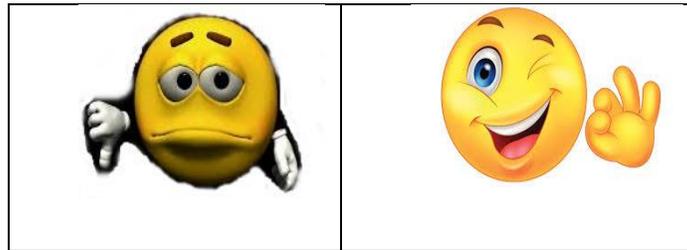

*Figure 2: Visual stimuli in the main experiment: Two emojis indicating a positive (happy) or negative (sad/angry) emotional valance.*

**Statistical Analysis**

In the present study, the dependent variable was categorical, rather than continuous, as participants chose between two emojis. Hence, general linear models, such as ANOVA, which are not optimal for categorial dependent variables, could not be used (Jaeger, 2008). Furthermore, to adapt the procedure to children's attention span, a limited number (<25) of repetitions of each condition had to be used for each participant. This limitation does not allow using the central limit theorem (CRT) for normal distribution, which the ANOVA model assumes. Hence, data was analyzed using the Generalized linear mixed-effects model analyses (GLMM), an extension of LMM that allows categorical data analysis (Jaeger, 2008). This analytic technique is increasingly used for experimental psychology data (and elsewhere), as it offers advantages over the traditional ANOVA test (see Boisgontier & Cheval, 2016 for a detailed argumentation), and is well suited for handling categorical data without assuming normality.

With regard to the present study, the GLMM model has two main advantages. First, as noted, it is highly suitable for handling categorial dependent variables. Second, participants can be inserted as a random variable. Hence, the variance within a particular participant could be treated as a random effect, since this variance is not a result of the experimental manipulation (Dyke & Patterson, 1952). We therefore determined participants as a random factor and inserted the independent variables in a stepwise fashion. To determine the appropriate random structure of the model, we began (following Barr, Levy, Scheepers, & Tily, 2013) with the maximum model for Emoji data, including all fixed factors and their interactions, as well as a random intercept for participants and a by-participant random slope for each fixed factor.

The GLMM for binary data was fitted by using the Logit link function (Jaeger, 2008). The dependent variable was emoji selection (happy/sad). The within-participants predictors of tonal stability, probe note and timbre, and the between-participants predictors of age group and sex where used in the preliminary model. For variables with more than two levels, dummy variables were computed. For example, for the "tonal stability" variable, which has 3 levels (stable diatonic, unstable diatonic and chromatic), we took "unstable diatonic" as the reference level, and the two computed dummy variables were stable diatonic (e.g. the difference between unstable diatonic and stable diatonic), and chromatic (e.g. the difference between unstable diatonic and chromatic). In this example, a significant effect of the stable diatonic variable would mean that the level of the stable diatonic tone was significantly higher than that of unstable diatonic tone; similarly, a significant effect of the chromatic variable would mean that the level of the chromatic tone was significantly lower than that of the unstable diatonic tone. We chose the reference category to be the hypothesized middle category (i.e. in tonal stability the reference category was the unstable diatonic and the computed dummy variables were stable diatonic and chromatic). Note that for this reason the model intercept may not reach significant level, because it is calculated by the probability of the reference category of all variables to differ from random (0.5 in case of 1 being happy emoji and 0 the sad emoji). An interaction with each such variable means an occurrence of a significant simple effect with another variable. For instance, a significant interaction of the dummy variable "chromatic tone" with the age variable means that the difference between the unstable diatonic and the chromatic variables differs in the two age groups.

Finally, another variable was calculated from the averaged emoji selected per each participant, which determined how much the participant was biased to happy or sad selection, regardless of experimental condition. We started with a model including all main effects and some theoretically meaningful interactions. Here, we report t-values only for significant effects and interactions. Analyses were carried out using IBM SPSS statistical software. Since some of the null effects were important to validate, we also applied Bayesian paired t-tests using the JASP software.

**Results**

For a full description of variables and effects see table 1. For the tonal stability variable, two dummy variables were created: (1) The stable diatonic tone [1] compared to both unstable diatonic [6] and chromatic [4#] tones; (2) The chromatic tone compared to both unstable diatonic and stable diatonic tones (for descriptives, see Figure 3). The main effect of tonal stability was significant in both comparisons: the happier emoji was selected more often for the stable diatonic than for unstable diatonic and chromatic tones, t=4.12, *p*<.001; and the sadder emoji was selected more often for the chromatic tone than for the unstable diatonic and stable diatonic tones, t=-2.211, *p*=.027.

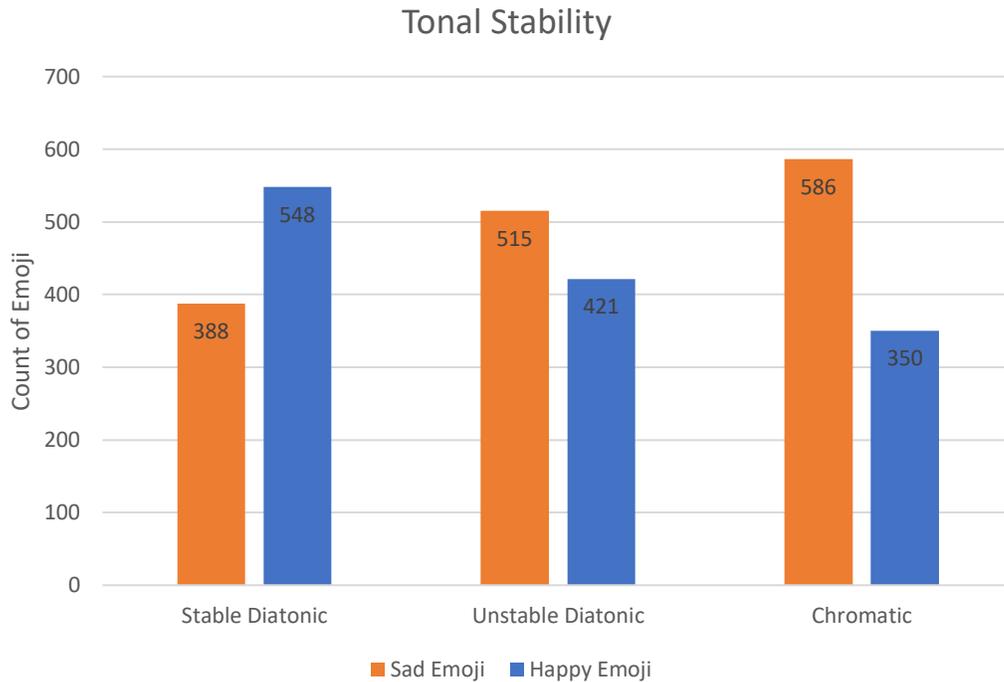

*Figure 3*: Total count of Emoji selection as a function of tonal stability categories, for Emoji (red bars) and sad Emojis (blue bars).

For the pitch height variable, two dummy variables were created: Eb4 (lowest) probe tones compared to both E4 and F4, and F4 (highest) probe tones compared to both E4 and Eb4 (for descriptives see Figure 4). The main effect of pitch height was significant for sadder emojis, selected for $E^b$ (the lowest probe tone) more than for E and F, t=4.155 *p*<.001. However, the difference between F (the highest probe) and E and $E^b$ did not reach significance (*p*>.05).

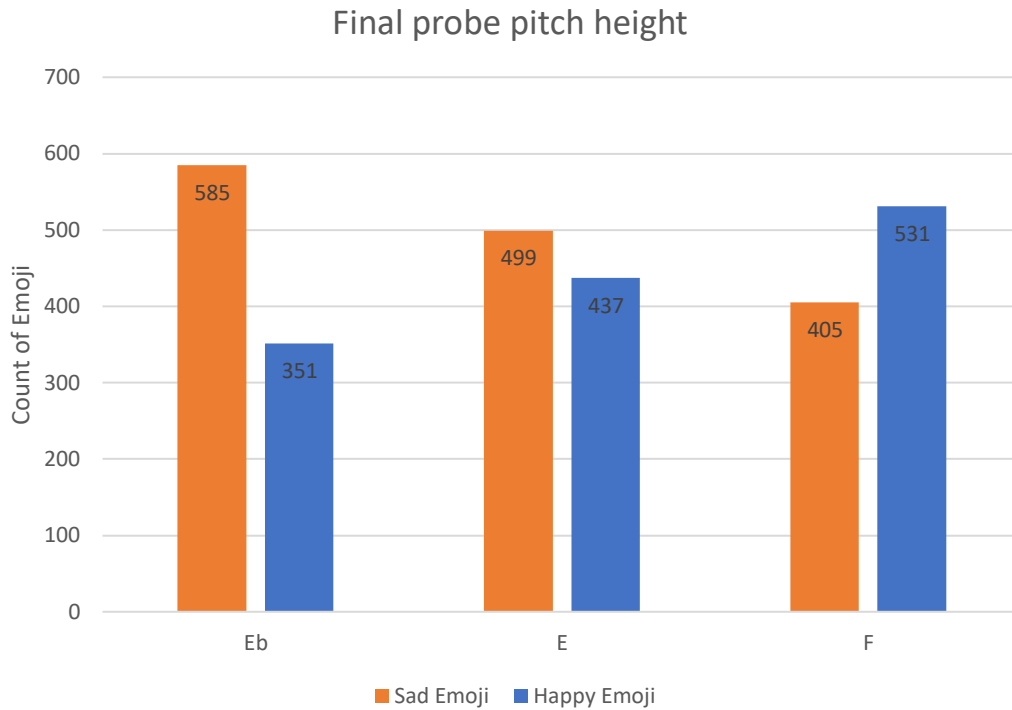

*Figure 4*:  *Total count of Emoji selection as a function of three probe-tone pitch height levels: E flat4, E4 and F4. Happy Emoji -- red bars; sad Emoji - blue bars.*

For the timbre variable, two dummy variables were created: piano compared to both guitar and woodwinds, and guitar compared to both woodwinds and piano (for descriptives see Figure 5). The main effect of timbre was significant for piano, as happier emoji selected more often for piano timbre than for guitar and woodwinds, t=3.09, *p*=.002. The difference between guitar compared to piano and woodwinds did not reach significance (*p*>.05). A post-hoc analysis was made to confirm a null effect of the difference between woodwinds and guitar timbers, using Bayesian model with Cauchy scale of .707 revealed that $BF_{01}$=5.685 and 5.661e-7% error rate.

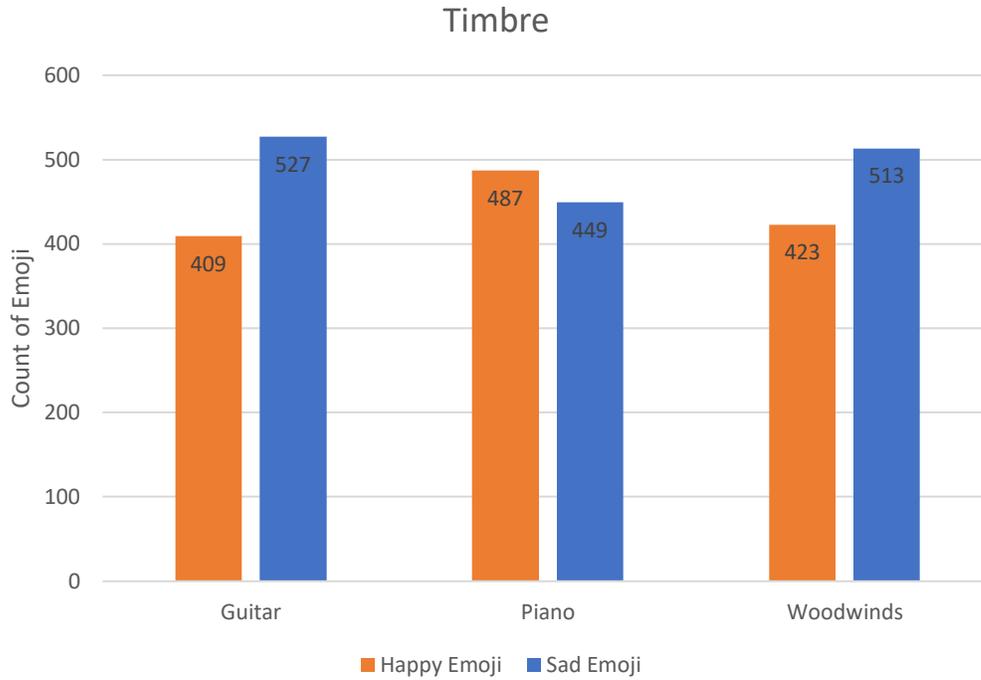

*Figure 5*:  Total count of Emoji selection as a function of three instrumental timbres (guitar, piano and woodwinds). Happy Emoji -- red bars; sad Emoji -- blue bars.

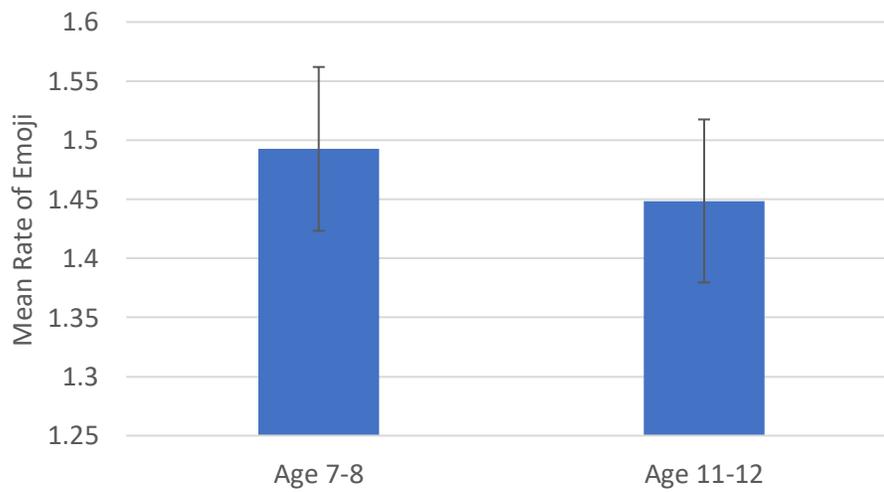

*Figure 6*: Mean Emoji ratings as a function age group. In Emoji ratings, 1=sad and 2=happy.

The main effect of age group was significant, indicating that, regardless of other variables, older children (ages 11-12) tended to select sadder emojis, compared to younger children (7-8 years old), t=-3.571, *p*<.001 (see Figure 6).

The interaction between age group and pitch height reached significance (see Figure 7). Older children (11 y.o), unlike younger ones, tended to select happier emoji when the probe tone was highest (F4), compared to the lower final probes, E4 and Eb4, t=4.546, *p*<.001.

All other main effects and interactions were not significant (all *p*s>.05).

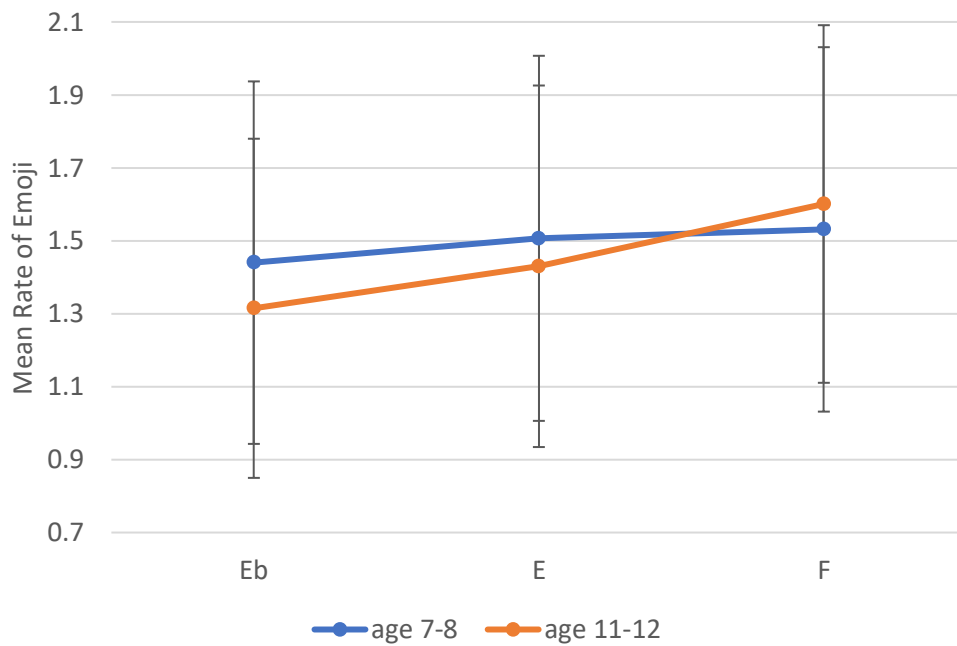

*Figure 7: Mean Emoji selection for age group 7-8 (blue) and age group 11-12 (red) as a function of final note pitch height (Eb, E and F).*

| Variable | Comparison | Coefficient | Std | t | Sig | 95% Confidence Interval: Lower | 95% Confidence Interval: Upper |
|---|---|---|---|---|---|---|---|
| ***Intercept*** | | -0.384 | 1.671 | -0.23 | 0.818 | -3.66 | 2.893 |
| ***Tonal Stability*** | Stable Diatonic > Unstable Diatonic | 0.511 | 0.124 | 4.12 | **<0.001** | 0.268 | 0.754 |
| | Chromatic < Unstable Diatonic | -0.225 | 0.102 | -2.211 | **0.027** | -0.425 | -0.025 |
| ***Pitch height*** | Low Final Probe (Eb) < middle Final Probe (E) | -0.405 | 0.097 | -4.155 | **<0.001** | -0.596 | -0.214 |
| | High Final Probe (F) > Middle Final Probe (E) | 0.144 | 0.149 | 0.966 | 0.334 | -0.148 | 0.435 |
| ***Instrumental timbre*** | Piano > Woodwinds | 0.3 | 0.097 | 3.09 | **0.002** | 0.109 | 0.49 |
| | Guitar < Woodwinds | -0.066 | 0.097 | -0.68 | 0.496 | -0.257 | 0.124 |
| ***Age*** | Older Age Group < Younger Age Group | -0.352 | 0.098 | -3.571 | **<0.001** | -0.545 | -0.159 |
| | Older Age Group*High Final note (F) | 0.764 | 0.168 | 4.546 | **<0.001** | 0.435 | 1.094 |
| ***Sex*** | Male > Female | -0.099 | 0.115 | -0.857 | 0.391 | -0.324 | 0.127 |
| ***Optimism*** | High Optimism | 0.518 | 0.083 | 6.267 | **<0.001** | 0.356 | 0.68 |

Table 1: output for the fixed effects from the generalized linear mixed-effects model used in the main experiment with tonal stability, pitch height, instrumental timbre, age group and gender as fixed variables. Note that the intercept is calculated by the hypothesized middle reference category of the dependent variables, namely unstable diatonic, middle final probe (E), and woodwinds for tonal stability, pitch height and instrumental timbre respectively.

## Discussion

Two main goals were set in this study. First, we investigated whether children associate tonal stability with emotional valence, particularly when tonal closure is implied. While comparable associations were found for adults, they have not yet been examined in children. We therefore investigated children's emotional associations of tonality in a developmental context, comparing two age groups (seven and 11 years old).

Second, we aimed to examine systematically how tonality interacts with other central musical dimensions (pitch height and instrumental timbre) in conveying emotion – whether emotional effects of tonality are independent of those of other musical features. That goal comprises both material and methodological aspects: while specifically addressing the interaction of tonality with pitch height and timbre in children's perception of musical emotion, we also sought, more generally, to demonstrate how the experimental design introduced here could facilitate a systematic examination of the ways tonality interacts with other musical dimensions. As the novel "reversed probe-tone" paradigm introduced here is particularly adapted to children, it could enable further examinations of such interactions in a larger developmental study, comprising both adults and children of diverse ages (including younger children).

Our findings suggest that children ages 7-11 indeed robustly associate tonal stability and closure with emotional valence, thus corroborating developmental models (McPherson & Schubert, 2016) which suggest that around this age children's emotional connotations of music are primarily based on schematic processes, deducing emotional expression from structural features. When tonal priming implying closure is introduced, stable probe tones (realizing closure expectations) were associated by children with positive emotional valence, while unstable tones, denying closure expectations, were associated with negative valence.

Importantly, these effects were secure in children as early as 7 years of age and did not significantly differ between younger (7 y.o.) and older (11 y. o.). The effect of tonality on perceived emotion was age-independent even though the age-group variable itself produced a significant main effect (younger children tended to associate the musical stimuli generally with more positive valence, compared to older children; Figure 6) and significantly interacted with musical features other than tonal stability (older children, but not younger ones, associated higher pitch with positive valence; Figure 7). The effect of tonal stability on perceived emotion was also consistent for different pitch heights and for different instruments, as neither of these variables significantly interacted with tonal stability. Importantly, that was the case though both variables themselves affected participants' responses, producing significant main effects: probe tones lower in pitch were associated with negative valence, compared to higher probes (Figure 4), and piano sounds, relative to guitar and woodwinds, were associated with positive valence (Figure 5). In sum, this study suggests that children associate levels of tonal closure with emotional valence, and that this this association is robust and consistent across age groups and musical attributes.

**Qualifications and suggestions for further studies**

This investigation of children's emotional associations to tonal closure is only a first, limited step. Tonal progressions continuously generate an evolving set of predictions, involving several levels and types of hierarchic structures (Lerdahl & Jackndoff, 1999). Each of these, alone and in interaction with other musical dimensions, such as melodic contour, or rhythm and meter (Meyer, 1956; Narmour, 1990) may carry its own emotional connotations – evidently more subtle than the dichotomic notion of valence presented here. Even in a simple diatonic setting, changes in harmonic context may substantially alter the emotional connotations of a given scale degree (Arthur, 2018). This complexity and subtlety may be what empowers tonal syntax as a carrier of emotion, and it calls for further exploration of that intriguing model of the mind – an exploration that could be facilitated using the paradigm introduced here.

Further limitations of this study should be addressed in further work. To facilitate the task and adapt it to children's capacities, we used only three different scale degrees (each representing a different level of tonal closure) as probe tones. Further studies could provide a more complete picture of children's mapping of tonal hierarchy onto emotional valence by using the entire gamut of 12 pitch-classes. The interaction of mode (major vs. minor) with tonal stability is also an important issue for future studies, as mode has been shown to be associated by children with emotional valence (Gregory, Worrall & Sarge, 1996; Dalla Bella et al., 2001; Ziv & Goshen, 2006; Nieminan et al, 2012).[2] Furthermore, investigating the emotional associations of tonal stability while utilizing a wider gamut of emotional dimensions or categories (beyond the simple valence dichotomy presented here) is certainly worth exploring.

**Conclusions**

While children's implicit perception of tonality has itself been amply studied, the emotional connotations of their tonal perception were not. This initial attempt of closing that gap provides a behavioral evidence that children (7 years and older) are sensitive to emotional cues suggested by tonal closure and its denial. This finding suggests that musical syntax, particularly structures denoting types and degrees of closure, are perceived by children not as mere abstractions, but as correlates of emotional states. Our experimental paradigm, adapted to younger children, could enable further investigation of children's emotional associations of tonality and their interactions with effects of other musical features.

---

[2] . A replication of this study with minor mode stimuli was unfortunately hindered by the COVID-19 crisis.

**Acknowledgment / Funding**

Research was supported by Israel Science Foundation (ISF) Grant 1920/16 to Zohar Eitan.

**References**

Arthur, C. (2018). A Perceptual study of scale-degree qualia in context. *Music Perception: An Interdisciplinary Journal, 35*(3), 295-314.

Adachi, M., & Trehub, S. E. (1998). Children's expression of emotion in song. *Psychology of Music, 26*(2), 133-153.

Barr, D. J., Levy, R., Scheepers, C., & Tily, H. J. (2013). Random effects structure for confirmatory hypothesis testing: Keep it maximal. *Journal of memory and language, 68*(3), 255-278.

Boisgontier, M. P., & Cheval, B. (2016). The ANOVA to mixed model transition. *Neuroscience & Biobehavioral Reviews, 68*, 1004-1005.

Carlsen, J. C. (1981). Some factors which influence melodic expectancy. *Psychomusicology: A Journal of Research in Music Cognition, 1*(1), 12.

Christensen, P., & James, A. (2008). Introduction: Researching children and childhood: Cultures of communication. In P. Christensen, & A. James (Eds.) *Research with children: Perspectives and practices* (2nd ed.) (pp. 1-9). Taylor & Francis.

Cohen, A. J. (2000). Development of tonality induction: Plasticity, exposure, and training. *Music Perception: An Interdisciplinary Journal, 17*(4), 437-459.

Costa-Giomi, E. (2003). Young children's harmonic perception. In G. Avanzini, C. Faienze, D. Minciacchi, L. Lopez, & M. Majno (Eds.), *The neurosciences and music: Annals of the New York Academy of Sciences* (Vol. 999, pp. 477 – 484). New York: New York Academy of Sciences.

Corrigall, K. A., & Schellenberg, G. E. (2016). Music cognition in childhood. In G. E. McPherson (Ed.), *The child as a musician: A handbook of musical development* (2nd ed.) (pp. 81-101). Oxford: Oxford University Press.

Corrigall, K. A., & Trainor, L. J. (2019). Electrophysiological Correlates of Key and Harmony Processing in 3-year-old Children. *Music Perception: An Interdisciplinary Journal, 36*(5), 435-447.

Cuddy, L. L., & Badertscher, B. (1987). Recovery of tonal hierarchy: some comparisons across age and level of musical experience. *Perception and Psychophysics*, *41*, 609-620.

Cuddy, L. L., & Lunney, C. A. (1995). Expectancies generated by melodic intervals: Perceptual judgments of melodic continuity. *Perception & Psychophysics, 57*(4), 451-462.

Dalla Bella, S., Peretz, I., Rousseau, L., & Gosselin, N. (2001). A developmental study of the affective value of tempo and mode in music. *Cognition*, *80*(3), B1-B10.

G.V. Dyke, H.D. Patterson (1952), Analysis of factorial arrangements when the data are proportions Biometrics, 8, pp. 1-12

Dowling, W. J. (1999). The development of music perception and cognition. In D. Deutcsh (Ed.), *The psychology of music* (2nd ed.) (pp. 603-625). San Diego: Academic Press.

Eerola, T., Ferrer, R., & Alluri, V. (2012). Timbre and affect dimensions: Evidence from affect and similarity ratings and acoustic correlates of isolated instrument sounds. *Music Perception: An Interdisciplinary Journal, 30*(1), 49-70.

Eerola, T., & Vuoskoski, J. K. (2013). A review of music and emotion studies: Approaches, emotion models, and stimuli. *Music Perception: An Interdisciplinary Journal, 30*(3), 307-340.

Eitan, Z., & Timmers, R. (2010). Beethoven's last piano sonata and those who follow crocodiles: Cross-domain mappings of auditory pitch in a musical context. *Cognition*, 114, 405–422.

Fétis, F. J. (1844). *Traité complet de la théorie et de la pratique de l'harmonie*. Brussels: Conservatoire de Musique; Paris: Maurice Schlesinger. English edition, as *Complete Teatiste on Theory and Practice of Harmony,* Translated by Peter M. Landey. Harmonologia: Studies in Music Theory 13. Hillsdale, N.Y.: Pendragon Press, 2008.

Ferrer, R., & Eerola, T. (2011). Semantic structures of timbre emerging from social and acoustic descriptions of music. *EURASIP Journal on Audio, Speech, and Music Processing*, *2011*(1), 11.

Gabrielsson, A. (2001). Emotion perceived and emotion felt: Same or different? *Musicae Scientiae*, *5* (1_suppl), 123-147.

Gabrielsson, A. (2011). The relationship between musical structure and perceived expression. Hallam, S., Cross, I., & Thaut, M. (Eds.). (2011). *Oxford handbook of music psychology* (pp. 141-150). Oxford University Press.

Gabrielsson, A., & Juslin, P. N. (1996). Emotional expression in music performance: Between the performer's intention and the listener's experience. *Psychology of Music, 24*(1), 68-91.

Gabrielsson, A., & Lindstrom, E. (2010). The role of structure in the musical expression of emotions. In P. N. Juslin, & J. A. Sloboda (Eds.), *Music and emotion: Theory, research, applications* (pp. 367-399). Oxford University Press.

Gebauer, L., Kringelbach, M. L., & Vuust, P. (2015). Predictive coding links perception, action, and learning to emotions in music: Comment on "The quartet theory of human emotions: An integrative and neurofunctional model" by S. Koelsch et al. *Physics of life reviews*, *13*, 50-52.

Granot, R., & Donchin, E. (2002). Do Re Mi Fa Sol La Ti——Constraints, Congruity, and Musical Training: An Event-Related Brain Potentials Study of Musical Expectancies. *Music Perception: An Interdisciplinary Journal, 19*(4), 487-528.

Gregory, A. H., Worrall, L., & Sarge, A. (1996). The development of emotional responses to music in young children. *Motivation and Emotion, 20*(4), 341-348.

Huron, D. (2006). *Sweet anticipation: Music and the psychology of expectation.* Cambridge, MA: MIT Press.

Jaeger, T. F. (2008). Categorical data analysis: Away from ANOVAs (transformation or not) and towards logit mixed models. *Journal of memory and language, 59*(4), 434-446.

James, W. (1894). The physical basis of emotion. *Psychological Review*, 7, 516-529.

Juslin, P. N., Harmat, L., & Eerola, T. (2014). What makes music emotionally significant? Exploring the underlying mechanisms. *Psychology of Music*, *42*(4), 599-623.

Juslin, P. N., & Västfjäll, D. (2008). Emotional responses to music: The need to consider underlying mechanisms. *Behavioral and brain sciences*, *31*(5), 559-575.

Kragness, H. E., & Trainor, L. J. (2018). Young children pause on phrase boundaries in self-paced music listening: The role of harmonic cues. *Developmental psychology, 54*, 842-856.

Koelsch, S. (2011). Toward a neural basis of music perception–a review and updated model. *Frontiers in psychology*, *2*, 110.

Koelsch, S., Fritz, T. & Schlaug, G. (2008). Amygdala activity can be modulated by unexpected chord functions during music listening. *Neuroreport* 19, 1815–1819.

Koelsch, S., Vuust, P., & Friston, K. (2018). Predictive processes and the peculiar case of music. *Trends in Cognitive Sciences*.

Krumhansl, C. L., & Keil, F. C. (1982). Acquisition of the hierarchy of tonal functions in music. *Memory & cognition, 10*(3), 243-251.

Krumhansl, C. L., & Kessler, E. J. (1982). Tracing the dynamic changes in perceived tonal organization in spatial representation of musical keys. *Psychological Review, 89*(4), 334-368.

Krumhansl, C. L. (1990). *Cognitive foundations of musical pitch*. New York: Oxford University Press.

Krumhansl, C. L. (2004). The cognition of tonality- as we know it today. *Journal of New Music Research,* 33, 253-268.

Krumhansl, C. L., & Cuddy, L. L. (2010). The theory of tonal hierarchies in music. In M. R. Jones, R. R. Fay, & A. N. Popper (Eds.), *Music Perception* (pp. 51-87). New York: Springer.

Lamont, A., & Cross, I. (1994). Children's cognitive representations of musical pitch. *Music Perception: An Interdisciplinary Journal, 12*(1), 27-55.

Lerdahl, F., and Jackendoff, R. (1999). *A Generative Theory of Music*. Cambridge: MIT Press.

Lundqvist, L. O., Carlsson, F., Hilmersson, P., & Juslin, P. N. (2009). Emotional responses to music: Experience, expression, and physiology. *Psychology of music*, *37*(1), 61-90.

McPherson, G. E., & Schubert, E. (2016) Underlying mechanisms and processes in the development of emotion perception in music. In G. McPherson (Ed.), *The child as musician: A handbook of musical development* (2nd ed.) (pp. 221–243). Oxford: Oxford University Press.

Maimon, N., Lami, D., & Eitan, Z. (2018). Emotional Associations to the Western Tonal System. Poster presentation at the 5th Israel Conference on Cognition Research (ISCOP), Akko (Israel), February 2018.

Maimon, N., Lami, D., & Eitan, Z. (under review). Cross-modal correspondence between tonal hierarchy and visual brightness: Associating syntactic structure and perceptual dimensions across modalities.

McPherson, G. E., & Schubert, E. (2016) Underlying mechanisms and processes in the development of emotion perception in music. In G. McPherson (Ed.), *The child as musician: A handbook of musical development* (2nd ed.) (pp. 221–243). Oxford: Oxford University Press.

Meyer, L. (1956). *Emotion and meaning in music.* Chicago, IL: Chicago University Press.

Miceli, M., & Castelfranchi, C. (2014). *Expectancy and emotion*. Oxford: Oxford University Press.

Müller, U., Carpendale, J. I., & Smith, L. (Eds.). (2009). *The Cambridge companion to Piaget*. Cambridge University Press.

Narmour, E. (1990). *The analysis and cognition of basic melodic structures: The Implication-Realization Model.* University of Chicago Press.

Ondobaka, S., Kilner, J., & Friston, K. (2017). The role of interoceptive inference in theory of mind. *Brain and cognition*, *112*, 64-68.

Pearce, M. T. (2018). Statistical learning and probabilistic prediction in music cognition: mechanisms of stylistic enculturation. *Annals of the New York Academy of Sciences*.

Pereira, M. R., Barbosa, F., de Haan, M., & Ferreira-Santos, F. (2019). Understanding the development of face and emotion processing under a predictive processing framework. *Developmental psychology*, *55*(9), 1868

Proulx, T., Inzlicht, M., & Harmon-Jones, E. (2012). Understanding all inconsistency compensation as a palliative response to violated expectations. *Trends in cognitive sciences*, *16*(5), 285-291.

Rameau, J. P. (1722). *Traité de l'harmonie reduite à ses principes naturels*. Paris: Ballard.

Schellenberg, E. G., Bigand, E., Poulin-Charronnat, B., Garnier, C., & Stevens, C. (2005). Children's implicit knowledge of harmony in Western music. *Developmental Science, 8*(6), 551-566.

Rao, R. P., & Ballard, D. H. (1999). Predictive coding in the visual cortex: a functional interpretation of some extra-classical receptive-field effects. *Nature neuroscience*, *2*(1), 79.

Schoenberg, A. (1978). *Theory of Harmony*, trans. Roy E. Carter (Berkeley: University of California Press, 1978), 25.

Seth, A. K. (2013). Interoceptive inference, emotion, and the embodied self. *Trends in cognitive sciences*, *17*(11), 565-573.

Shanahan, D. (2017). Musical Structure: Tonality, Melody, Harmonicity, and Counterpoint. In R. Ashley, & R. Timmers (Eds.), *The routledge companion to music cognition* (pp. 129-140). Taylor & Francis.

Shepard, R. N. (1964). Circularity in judgments of relative pitch. *The Journal of the Acoustical Society of America, 36*(12), 2346-2353.

Sloboda, J. A. (1985). *The musical mind: The cognitive psychology of music.* Ney York: Oxford University Press.

Sloboda, J. A. (1991). Music structure and emotional response: Some empirical findings. *Psychology of music*, *19*(2), 110-12

Speer, J. R., & Meeks, P. U. (1985). School children's perception of pitch in music. *Psychomusicology: A Journal of Research in Music Cognition, 5*, 46-56.

Steinbeis, N., Koelsch, S., & Sloboda, J. A. (2006). The role of harmonic expectancy violations in musical emotions: Evidence from subjective, physiological, and neural responses. *Journal of cognitive neuroscience*, *18*(8), 1380-1393.

Thompson, W. F. (2015). *Music, thought, and feeling: Understanding the psychology of music*. Oxford university press.

Tillmann, B. (2005). Implicit investigations of tonal knowledge in nonmusician listeners. *Annals of the New York Academy of Sciences*, *1060*(1), 100-110.

Tillmann, B., & Bigand, E. (2004). The relative importance of local and global structures in music perception. *The Journal of Aesthetics and Art Criticism, 62*(2), 211-222.

Trainor, L. J. (2005). Are there critical periods for musical development? Developmental Psychobiology: *The Journal of the International Society for Developmental Psychobiology, 46*(3), 262-278.

Trainor, L. J. (2012). Predictive information processing is a fundamental learning mechanism present in early development: evidence from infants. *International Journal of Psychophysiology*, *83*(2), 256-258.

Trainor, L. J., & Trehub, S. E. (1994). Key membership and implied harmony in Western tonal music: Developmental perspectives. *Perception & Psychophysics, 56*(2), 125-132.

Trainor, L. J., Tsang, C. D., & Cheung, V. H. (2002). Preference for sensory consonance in 2-and 4-month-old infants. *Music Perception: An Interdisciplinary Journal, 20*(2), 187-194.

Trainor, L. J., & Corrigall, K. A. (2010). Music acquisition and effects of musical experience. In M. R. Jones, R. R. Fay, & A. N. Popper (Eds.), *Music Perception* (pp. 89-127). New York: Springer.

Trehub, S. E. (1987). Infants' perception of musical patterns. *Perception & Psychophysics, 41*(6), 635-641.

Trehub, S. E., Weiss, M. W. (2017). Music cognition: Developmental and Multimodal Perspectives. In R. Ashley, & R. Timmers (Eds.), *The routledge companion to music cognition* (pp. 403-414). Taylor & Francis.

Wilson, S. J., Wales, R. J., & Pattison, P. (1997). The representation of Tonality and Meter in Children Aged 7 and 9. *Journal of Experimental Child Psychology, (64)*, 42-66.

Yin, L., Chen, X., Sun, Y., Worm, T., & Reale, M. (2008, September). *A high-resolution 3D dynamic facial expression database*. In Automatic Face & Gesture Recognition. FG'08. Eighth IEEE International Conference on. (pp. 1-6) IEEE.

Zimmerman, M. (1993). An overview of developmental research in music. *Bulletin of the Council for Research in Music Education*, (116), 1-21.

Ziv, N., & Goshen, M. (2006). The effect of "sad" and "happy" background music on the interpretation of a story in 5- to 6-year-old children. *British Journal of Music Education*, 23, 303-314.

## *Appendix: Pretest 1*

**Method**

**Participants.** 21 children aged 7 and 11 served as participants: 13 (eight females, five males) in the younger group, eight (five females, three males) in the older group. All the 11-year-olds had a history of 6 to 24 months of music lessons, while six of them played a musical instrument for 6 to 24 months. Five of the 7-year-old had 6 to 12 months of Western music lessons but none of them played a musical instrument. All were white native Hebrew speakers, living in a small middle-class community. Children were sampled from a non-profit after-school community center. All had everyday exposure to various genres of Western tonal music (e.g. pop, rock etc.). Parent's children signed informed consent according to declaration of Helsinki. Children agreed to participate willingly and could quit at any time with no sanction. The children received a small sweet prize as a compensation.

**Apparatus.** The stimuli were presented using two computers: a Dell N5110 computer with a 15.6 inches screen, and a Dell E6420 computer with a 14 inches screen. Auditory stimuli were delivered through Behringer HPM1000 headphones. The experiment was programmed and performed using Matlab. The participants (up to 5 children) sat in one room in front of separate computers, distant from each other and without eye contact.

**Auditory stimuli:** Since pitch height is known to be associated with valence (Eitan & Timmers, 2010; Gabrielsson & Lindstrom, 2010), all auditory stimuli were created using Shepard tones (Shepard, 1964) to minimize such pitch-height effects. Shepard tones are tones with a specific pitch chroma (pitch note), sounded in 5 octaves simultaneously. A loudness envelope across the frequency range of 77.8-2349 Hz was used, with gradual increase/decrease at its endings (Shepard, 1964). This loudness envelope generates tones with a clear pitch chroma but an ambiguous pitch height (i.e., the register in which a pitch is perceived – whether it is perceived in a higher or lower octave – is ambiguous and determined contextually). All stimuli were sounded in 71 dB.

*Table 1 (app.). Musical elements used in pretest 1*

| Element | Composition |
| --- | --- |
| G major chord | G B D |
| D-flat Major chord | $D^b$ F $A^b$ |
| G minor chord | G $B^b$ D |
| D-flat minor chord | $D^b$ $F^b$ $A^b$ |
| IV-V-I cadence in G major | C major chord, D major chord, G major chord |
| IV-V-I cadence in D-flat major | $G^b$ major chord, $A^b$ major chord, $D^b$ major chord |
| IV-V-I cadence in G minor | C minor chord, D major chord, G minor chord |
| IV-V-I cadence in D-flat minor | $G^b$ minor chord, $A^b$ major chord, $D^b$ minor chord |

Each trial consisted of a key-establishing element followed by a probe. There were eight possible elements (see Table 1): two element types (a minor or major triad, presumably perceived as a tonic chord, and a IV-V-I cadence) in four keys (G major, G minor, D-Flat major and D-Flat minor), and 12 possible probes -- the 12 tones of the chromatic scale (12 notes from C to B). The triad (single chord) element was played for 500 ms; in the cadence element, each chord was played for 500 ms, followed by a 2500 ms silence. Then, following a silence of 1000 ms, a probe tone was played for 500 ms.

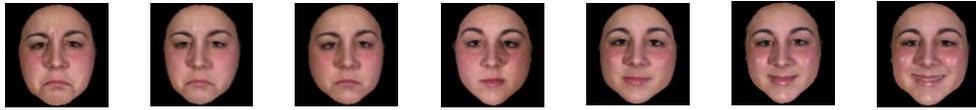

*Figure 1 (app.): One of four gradient scales of emotional faces (source: the BU-3DFE Database).*

**Visual stimuli:** The visual stimuli consisted of seven photographed faces, their emotional valence gradually ranging from negative (sad) to positive (happy). We used photographs of veridical faces selected from the BU-3DFE Database (Yin et al., 2008). In the present experiment, we selected four Caucasian individuals (two men and two women) and generated four 7-level continuums by selecting the first three levels of sadness and the first three levels of happiness and adding the neutral expression as the middle level (see Figure 2).

**Procedure.** Ethics approval was obtained from the Levinsky College Ethics Committee. A signed informed consent was obtained from parents of all participants.

Participants first heard a recorded explanation of the task, with an identical printed text simultaneously presented on the screen (both in Hebrew). For each block, one facial continuum out of the four appeared on the screen throughout the entire block. For each participant, a single graded ordering of stimuli – happier faces to the right or happy faces to the left – was used throughout the experiment, these orderings counterbalanced among participants in each group (younger children; older children). In each trial, after listening to the element followed by the probe, participants were asked to mark the face which, according to their subjective judgment, best matched the expression evoked by the probe tone. The auditory and visual stimuli were presented simultaneously with the instruction "Press the face describing the 'feeling' of the last note, as it relates to notes preceding it." Participants were encouraged to use the full range of pictures. After each block, participants were asked about their confidence in the judgments they had provided. The text "How confident were you with the ratings of the last block?" appeared on the screen, and participants provided a 1-7 rating by pressing the appropriate numeral on the upper row of the keyboard.

The experiment consisted of 10 blocks. The first two blocks served as practice. The elements for these blocks were quasi-randomly drawn from the eight possible elements, and consisted of either a cadence or a triad, in either G or D Flat (counterbalanced across subjects). They were followed by eight experimental blocks, one for each of the eight possible elements, randomly mixed. The same element was used throughout any given block of trials. Each block consisted of 14 consecutive trials. The first two trials served as practice, using two randomly selected probes. In the following 12 experimental trials, the 12 chromatic tones, randomly mixed, were used as probes. In sum, the data of each participant included 96 (8X12) trials. Ten seconds of white noise (71db) were inserted between the blocks A virtual illustration of a gold bag was shown after the $4^{th}$, $7^{th}$ and $10^{th}$ block (the participants accumulated three virtual gold bags in order to encourage persistence). The experiment lasted approximately 30 minutes.

**Results**

A repeated-measures analysis of variance (ANOVA) was used, with age (seven years olds vs. 11 years olds) as a between-participants variable and tonal stability and mode (major/minor) as within-participant variables. The tonal stability variable included three categories: *Stable diatonic* (the tonic chord members – the $1^{st}$, $3^{rd}$, and $5^{th}$ degrees of the scale; that is, tones 1, 5, 8 of the 12 chromatic tones in major and 1, 4, 8 in minor), *Unstable diatonic* ($2^{nd}$, $4^{th}$, $6^{th}$ and $7^{th}$ degrees of the scale: 3, 6, 10, 12 of the 12 chromatic tones in major, and 3, 6, 7, 11, 12 in minor, where both options for the $7^{th}$ scale-degree, raised [leading-tone] and natural, were included) and *chromatic* *tones* (out-of-key notes: 2, 4, 7, 9, 11 of the 12 tones in major and 2 ,5 ,7, 10 in minor). For this variable we conducted planned comparisons, weighted in a repeated manner: stable diatonic>unstable diatonic>chromatic tones. For results summary, see Figure 3.

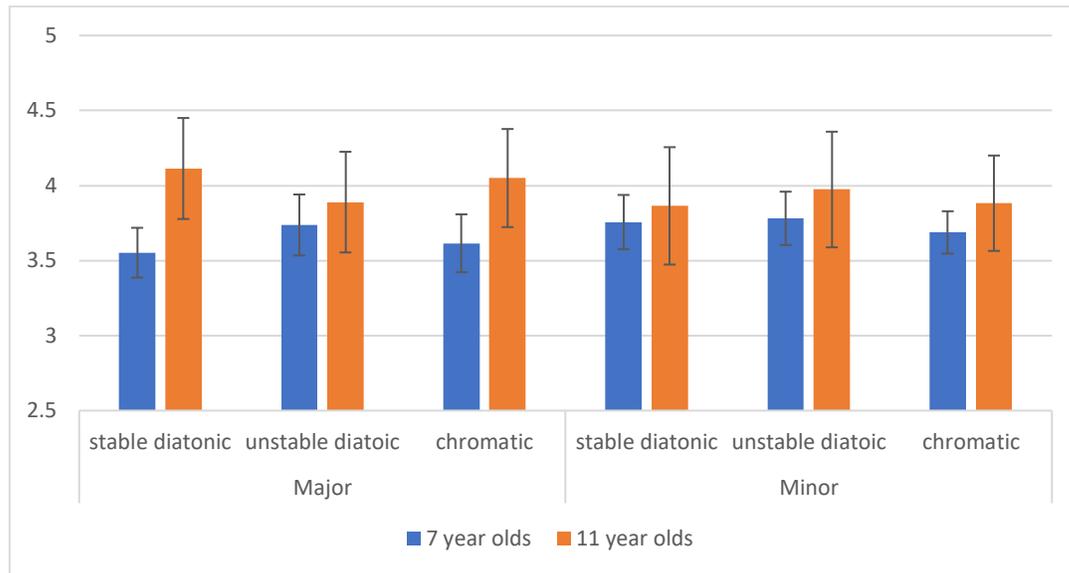

*Figure 2 (app.): Mean emotional faces matchings in Experiment 1 as a function of tonal stability categories, in major (right panel) and minor (left panel) modes and for 7-years old (blue bars) and 11-years old (red bars). 1=happiest face, 7=saddest face.*

The main effect of tonal stability was not significant, F<1. Planned comparisons revealed that emotional faces matchings did not significantly differ between any pair of categories (all Fs<1). There were no other significant main effects or interactions (all *p*s>0.05).

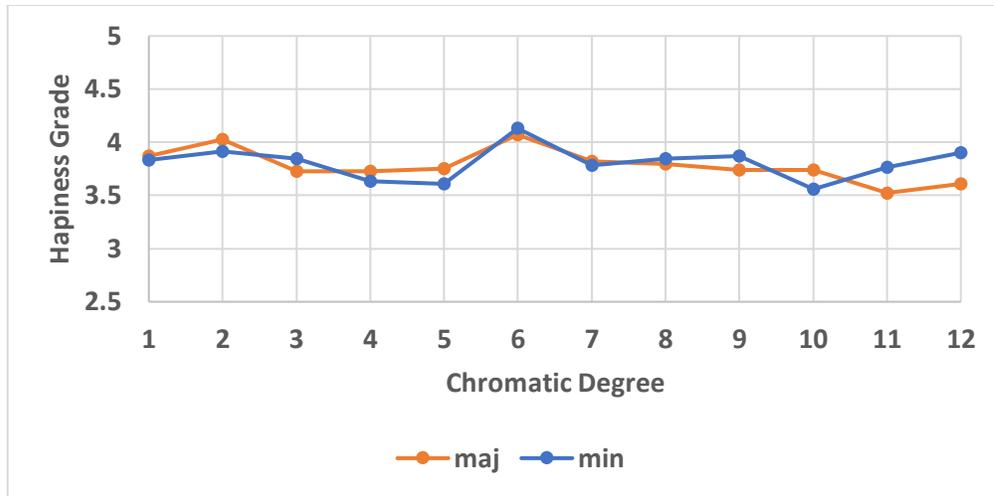

*Figure 3 (app.): Mean faces matchings in Major (red) and Minor (blue), in Experiment 1a as a function of scale degrees (1=tonic; 12=leading tone). 1=saddest face, 7=happiest face.*

As an exploratory analysis, we also examined whether the children exhibited any scale-degree related effects – even effects different from the tonal stability effects we predicted. To that purpose, we created a variable of "chromatic degree," -- the tonal chromatic degree of each probe, consisting 12 levels, with the tonic note designated 1 (e.g. the pitch-class G was designated as 1 in G Major and as 7 in C# Major). We conducted a second ANOVA with age (7-years old vs. 11-years old) as a between-participants variable and chromatic degree (1-12) and mode (major/minor) as within-participant variables. The main effect of chromatic degree was not significant, $F<1$, $p>.05$, that is, the emotional faces matchings of the different chromatic degrees were not significantly different. There was no significant interaction of chromatic degree with any of the other variables, and no other effects reached significance level (all $p$s>.05). For results summary see Figure 4.

*Confidence ratings*

The overall mean confidence rating was M=5.642 (SD=1.12). An ANOVA with age (7/11) as a between-participants variable and mode (major/minor) as a within–participant variable was conducted on the mean confidence ratings obtained. There were no main effects of age, M=5.644,

SD=1.44 for 7 years old and M=5.641, SD=1.216 for 11 years old, F(1,19)<1, or mode M=5.619, SD=1.344 for major mode and M=5.66, SD=1.374 for minor mode, F(1,19)<1, and no interaction between the two variables, F(1,19)=3.446, *p*=.079.